\documentclass[letterpaper,9pt,conference]{IEEEtran}
\IEEEoverridecommandlockouts
\usepackage{cite}
\usepackage{amsmath,amssymb,amsfonts}
\usepackage{algorithmic}
\usepackage{graphicx}
\usepackage{textcomp}
\usepackage{xcolor}
\DeclareMathOperator*{\argmax}{arg\,max}
\def\BibTeX{{\rm B\kern-.05em{\sc i\kern-.025em b}\kern-.08em
    T\kern-.1667em\lower.7ex\hbox{E}\kern-.125emX}}
    
\usepackage{algorithm}
\usepackage{etoolbox}
\usepackage{subcaption}
\usepackage{multirow}
\usepackage{pifont}
\usepackage{url}
\newcommand\numberthis{\addtocounter{equation}{1}\tag{\theequation}}
\renewcommand{\vec}[1]{\mathbf{#1}}
\renewcommand\footnoterule{
\kern -3pt
\hrule width 0.5\linewidth height 0.5pt
\kern 2pt}
\usepackage[T1]{fontenc}
\usepackage[normalem]{ulem}
\begin{document}
\title{Multi-Objective Optimization of ReRAM Crossbars for Robust DNN Inferencing under Stochastic Noise\vspace{-6pt}\thanks{This work was supported in part by the US National Science Foundation (NSF) under grants CNS-1955196, CNS-1955353, IIS-1845922, OAC-1910213, and by the USA Army Research Office grant W911NF-17-1-0485. Biresh Kumar Joardar was also supported in part by NSF Grant \#2030859 to the Computing Research Association for the CIFellows
Project.}}
\author{Xiaoxuan Yang$^*$, Syrine Belakaria$^\dagger$, Biresh Kumar Joardar$^*$, Huanrui Yang$^*$,\\ Janardhan Rao Doppa$^\dagger$, Partha Pratim Pande$^\dagger$, Krishnendu Chakrabarty$^*$, Hai (Helen) Li$^*$ \\$^*$ Department of Electrical and Computer Engineering, Duke University, Durham, NC, USA,\\
$^\dagger$ School of Electrical Engineering \& Computer Science, Washington State University, Pullman, WA, USA\\
$^*$ \{xy92, bireshkumar.joardar, huanrui.yang, krish, hai.li\}@duke.edu, $^\dagger$ \{syrine.belakaria, jana.doppa, pande\}@wsu.edu
\vspace{-18pt}}
\maketitle
\begin{abstract}
Resistive random-access memory (ReRAM) is a promising technology for designing hardware accelerators for deep neural network (DNN) inferencing.
However, stochastic noise in ReRAM crossbars can degrade the DNN inferencing accuracy.
We propose the design and optimization of a high-performance, area-and energy-efficient ReRAM-based hardware accelerator to achieve robust DNN inferencing in the presence of stochastic noise.
We make two key technical contributions.
First, we propose a stochastic-noise-aware training method, referred to as ReSNA, to improve the accuracy of DNN inferencing on ReRAM crossbars with stochastic noise.
Second, we propose an information-theoretic algorithm, referred to as CF-MESMO, to identify the Pareto set of solutions to trade-off multiple objectives, including inferencing accuracy, area overhead, execution time, and energy consumption.
The main challenge in this context is that executing the ReSNA method to evaluate each candidate ReRAM design is prohibitive.
To address this challenge, we utilize the continuous-fidelity evaluation of ReRAM designs associated with prohibitive high computation cost by varying the number of training epochs to trade-off accuracy and cost.
CF-MESMO iteratively selects the candidate ReRAM design and fidelity pair that maximizes the information gained per unit computation cost about the optimal Pareto front.
Our experiments on benchmark DNNs show that the proposed algorithms efficiently uncover high-quality Pareto fronts.
On average, ReSNA achieves $2.57\%$ inferencing accuracy improvement for ResNet20 on the CIFAR-10 dataset with respect to the baseline configuration. Moreover, CF-MESMO algorithm achieves $90.91\%$ reduction in computation cost compared to the popular multi-objective optimization algorithm NSGA-II to reach the best solution from NSGA-II.
\end{abstract}
\section{Introduction}
Resistive random access memory~(ReRAM) has emerged as a promising nonvolatile memory technology due to its multi-level cell, small cell size, and low access time and energy consumption.
Prior work has shown that the crossbar structure of ReRAM arrays can efficiently execute matrix-vector multiplication~\cite{hu2012hardware,chen2020survey}, the predominant computational kernel associated with deep neural networks~(DNNs).
ReRAM-based accelerators for fast and efficient DNN training and inferencing have been extensively studied~\cite{chi2016prime,shafiee2016isaac,song2017pipelayer,long2018reram,fan2019red,yang2020retransformer}.

However, a key challenge in executing DNN inferencing \cite{EdgeAI-1,EdgeAI-2,TS-1} on ReRAM-based architecture arises due to nonidealities of ReRAM devices, which can degrade the accuracy of inferencing. Since DNN inferencing involves a sequence of forward computations over DNN layers, errors due to device nonidealities can propagate and accumulate, resulting in incorrect predictions. The nonidealities of ReRAM crossbars can be classified into two broad categories. The first category includes device defects (e.g., stuck-at-high or stuck-at-low resistance~\cite{chen2014rram}) and device reliability issues (e.g., retention failure~\cite{schechter2010use} and resistance drift~\cite{chen2013postcycling}) that are mostly deterministic in nature and have been addressed in prior work \cite{xu2015impact,liu2017rescuing,xia2017stuck,chen2017accelerator,li2019build,chakraborty2020geniex}. The second category includes stochastic noise in ReRAM devices that includes thermal noise~\cite{soudry2015memristor}, shot noise~\cite{feinberg2018making}, random telegraph noise (RTN)~\cite{ielmini2010resistance}, and programming noise~\cite{lee2012multi}. These nonidealities have not been studied for DNN inferencing in prior work.

This paper studies the impact of stochastic noise on DNN inferencing and shows that there is a significant degradation in inferencing accuracy due to the high amplitude of noise and reduced noise margin of high-resolution ReRAM cells. Prior algorithmic solutions~\cite{long2019design,joshi2020accurate} mitigate the accuracy degradation due to programming variations, but they are not effective in the presence of stochastic noise~\cite{he2019noise}. To overcome this challenge, we propose a ReRAM-based  Stochastic-Noise-Aware DNN training method~(ReSNA) that considers both hardware design configurations and stochastic noise. 

For DNN inferencing on ReRAM using ReSNA, key efficiency metrics include hardware area, execution time~(latency), and energy consumption. Therefore, we need to solve a complex multi-objective optimization (MOO) problem to achieve robust DNN inferencing on ReRAM crossbars. The input space consists of different ReRAM crossbar configurations, e.g., ReRAM cell resolution, crossbar size, temperature, and operational frequency. The output space consists of the accuracy of DNN inferencing and hardware efficiency metrics, e.g., hardware area, execution time, and energy consumption. The main challenge in solving this optimization problem is that the input space over ReRAM configurations contains a large number~(up to $10^7$) of available data points, and evaluation of each candidate ReRAM configuration involves executing the ReSNA method, which is computationally prohibitive~(e.g., it takes nearly $30$ GPU days to run the training on the crossbar simulator~\cite{he2019noise} for 100 configurations). Our goal is to efficiently uncover the Pareto optimal set of solutions representing the best possible trade-offs among multiple objectives. 

To solve this challenging MOO problem, we propose an information-theoretic algorithm referred to as Continuous-Fidelity Max-value Entropy Search for Multi-Objective Optimization (CF-MESMO). 
We formulate the continuous-fidelity evaluation by varying the number of training epochs for ReSNA to establish an appropriate trade-off between computation cost and accuracy.
In each MOO iteration, the candidate ReRAM design and fidelity (number of iterations of ReSNA training) pair is selected based on the maximization of the information gained per unit computation cost about the optimal Pareto front. We perform comprehensive experiments on benchmark DNNs and datasets to evaluate the proposed algorithms. Our results show that ReSNA can significantly increase DNN inferencing accuracy in the presence of stochastic noise on ReRAM crossbars, and CF-MESMO can achieve faster convergence and efficiently uncover high-quality Pareto fronts when compared to prior methods, including NSGA-II~\cite{deb2002fast} and a state-of-the-art single-fidelity multi-objective optimization method called MESMO~\cite{belakaria2019max}. 

The main contributions of this paper are as follows.
\begin{itemize}
\item Study of the impact of stochastic noise on DNN inferencing executed on ReRAM crossbars.
\item A hardware-aware training method, referred to as ReSNA, to overcome stochastic noise and improve DNN inferencing accuracy.
\item An efficient multi-objective optimization algorithm, referred to as CF-MESMO, to approximate optimal Pareto fronts in terms of inferencing accuracy and hardware efficiency.
\item Experimental results on a diverse set of benchmark DNNs and datasets to demonstrate the effectiveness of ReSNA and CF-MESMO and their superiority over state-of-the-art methods.
\end{itemize}

The remainder of this paper is organized as follows.
Section~\ref{sec2_related} discusses related prior work.
Section~\ref{sec3_problem} explains the problem setup, and Section~\ref{sec4_stochastic} highlights the impact of stochastic noise.
Section~\ref{sec5_training} presents the ReSNA approach, and Section~\ref{sec6_moo} presents the CF-MESMO algorithm.
Section~\ref{sec7_experiment} presents the experimental results.
Section~\ref{sec8_conclusion} concludes this paper.

\section{Related Prior Work}
\label{sec2_related}
We review related prior work on two key aspects of this paper---mitigating device stochastic noise for DNN inferencing and multi-objective optimization for hardware design.

There is limited prior work on mitigating the DNN inferencing accuracy loss due to stochastic noise. Yan et al.~\cite{yan2017closed} proposed a closed-loop circuit that utilizes the inferencing results to stabilize the DNN weights, but the effectiveness of this method was demonstrated only on small DNNs. Long et al.~\cite{long2019design} injected Gaussian noise during training to mimic programming noise, and Joshi et al.~\cite{joshi2020accurate} incorporated device programming variation extracted from experiments during training. However, these methods only considered programming noise while neglecting the other types of noise. Importantly, all these prior methods overlooked the impact of hardware configurations, such as the crossbar size and the resolution of the digital-to-analog converter (DAC) and the analog-to-digital converter (ADC).
%the crossbar size and 

He et al.~\cite{he2019noise} investigated the integration of stochastic noise during the training process, but their method failed to reach the desired DNN inferencing accuracy. Consequently, they suggested lowering the operational frequency such that the noise amplitude is low. In contrast to prior work, we consider all the four types of stochastic noise and propose a ReRAM hardware-aware training method to increase the inferencing accuracy even under high operational frequencies.

Considering both inferencing accuracy and hardware efficiency, we have a complex MOO problem for ReRAM-based hardware design.
%\noindent{\bf MOO algorithms for hardware design optimization.}
Candidate MOO algorithms for ReRAM design optimization can be classified into two broad categories. The first category of MOO algorithms has objective functions that are {\em cheap} to evaluate. AMOSA~\cite{bandyopadhyay2008simulated} and NSGA-II~\cite{deb2002fast} are two popular evolutionary algorithms that belong to this category. NSGA-II evaluates the objective functions for various combinations of input variables and organizes the candidate inputs into a hierarchy of subgroups based on the ordering of Pareto dominance. This method takes advantage of the similarity between members of each subgroup and the Pareto dominance and moves towards the promising area of the input space. Unfortunately, NSGA-II requires the evaluation of a large number of candidate inputs and is not suitable for our problem setting, where objectives are expensive. 

Second, for {\em expensive} objective functions, Bayesian optimization (BO) \cite{BO-Survey} is an effective framework. The key idea is to build a cheap statistical model from past function evaluations and use it to intelligently explore the input space for finding (near-)optimal solutions. Much of the prior work on BO is for single-objective optimization. There is limited work on multi-objective BO \cite{USEMO,ACDesign,belakaria2020PSD}, and MESMO \cite{belakaria2019max} is the state-of-the-art algorithm. In contrast to prior work, we exploit the continuous approximations of the objective functions (hardware-aware DNN training on ReRAM designs with varying the number of epochs) to minimize the computation cost to uncover high-quality Pareto front for hardware design. 

\section{Background and Problem Setup}
\label{sec3_problem}
In this section, we first explain how a trained DNN model is deployed on the ReRAM crossbars to perform inferencing. Subsequently, we describe the MOO problem to perform robust DNN inferencing using ReSNA. 
\subsection{DNN Inferencing on ReRAM Crossbars}
\begin{figure}[t]
\vspace{-12pt}
\centering
\includegraphics[width=
0.97\linewidth]{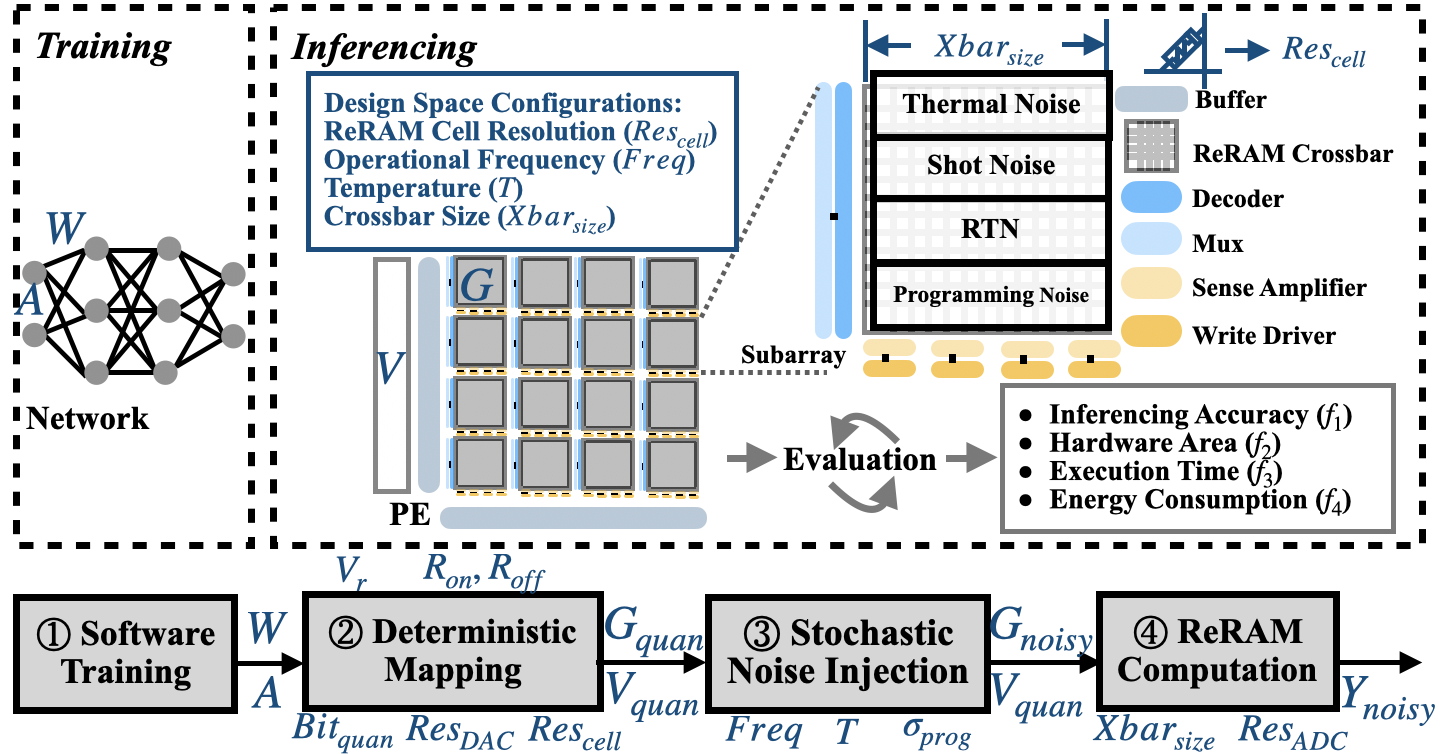}
\vspace{-6pt}
\caption{DNN inferencing process on ReRAM crossbars.}
\label{fig:map}
\end{figure}

Fig.~\ref{fig:map} illustrates the overall flow of deploying a trained DNN model on ReRAM crossbars for inferencing. There are four main steps, as explained below.
Table~\ref{table:parameter} summarizes the notation associated with the relevant parameters. 
\begin{table}[t]
\centering
\vspace{-6pt}
\caption{Parameters for inferencing on ReRAM crossbars.}
\label{table:parameter}
\vspace{-3pt}
\footnotesize
\begin{tabular}{|c||c|}
\hline
{\bf Notation} & {\bf Meaning} \\ \hline
$W$ &  DNN weight matrix     \\
$A$ &  Activations      \\ \hline
$Bit_{quan}$ & Bit number for quantization   \\
$R_{on},R_{off}$ & Low resistance and high resistance       \\
$Res_{cell}$ & ReRAM cell resolution      \\
$Res_{DAC},Res_{ADC}$ & DAC and ADC resolution      \\
$V_{r}$ & ReRAM read voltage      \\ 
$T$ & Temperature      \\
$Freq$ & Operational frequency      \\
$\sigma_{prog}$ & Programming noise standard deviation      \\
$Xbar_{size}$ & Crossbar size      \\ \hline
\end{tabular}
\vspace{-12pt}
\end{table}

%\vspace{0.25ex}
\noindent\ding{172}~\textit{Software training.}
For a given DNN architecture and training dataset, we first perform the training in software. The quantization-aware training technique~\cite{han2015deep_compression,zhou2016dorefa,yang2021bsq} can be used to quantize the activations and weights.

\vspace{0.25ex}

\noindent\ding{173}~\textit{Deterministic mapping.}
The objective of this step is to map the weight matrix of DNN $W$ and the set of activations $A$ to the conductance of the ReRAM cells $G_{quan}$ and crossbar input voltages $V_{quan}$, according to the resolutions of ReRAM devices and DACs, respectively.
When the ReRAM cell resolution is less than the number of bits used in quantization, i.e., $Res_{cell} < Bit_{quan}$, $\lceil{Bit_{quan}}/{Res_{cell}} \rceil$ cells are used to represent one weight. 
A kernel in a convolutional (Conv) layer needs to be first unrolled and mapped to a matrix.
As kernels are reused many times during convolution, the kernel in a Conv layer is typically duplicated and deployed on multiple crossbars. 
Therefore, multiple inputs can be processed simultaneously, increasing parallelism and improving throughput~\cite{song2017pipelayer}. 
For a fully-connected (FC) layer, each weight is associated with only one input neuron. Hence, duplication is not necessary. 
\vspace{0.25ex}

\noindent\ding{174}~\textit{Stochastic noise injection.} This step mimics the influence of stochastic noise on the conductance values.
The noise is modeled using probability distributions~\cite{feinberg2018making,he2019noise}. 
Here $G_{noisy}$ denotes the cell conductance in the presence of thermal noise, shot noise, RTN, and programming noise together.
Section~\ref{sec:noise} provides more details.

\vspace{0.25ex}

\noindent\ding{175}~\textit{ReRAM-based computation.}
This process accumulates the results obtained from ReRAM crossbars and employs ADCs to generate outputs.
Here $Y_{noisy}$ denotes the final output of DNN inferencing.

\vspace{0.25ex}

\noindent Note that the last three steps together emulate the deployment of DNN inferencing on ReRAM-based hardware. 
The deterministic mapping needs to be carried out only once.
Typically, we need to perform the third and fourth steps multiple times (e.g., ten times) to mimic multiple independent ReRAM deployments on the same device. Based on the multiple runs, we obtain an estimate of DNN inferencing accuracy. 

\subsection{MOO problem for ReRAM-based Designs}

Our goal is to find ReRAM-based designs with suitable DNN weights to optimize multiple objectives, including inferencing accuracy, hardware area, execution time, and energy consumption. 

\noindent {\bf ReRAM design space.} The ReRAM design configuration influences the output objectives.
For example, the parameters $Bit_{quan}$,~$Res_{DAC}$,~$Res_{ADC}$ listed in Table~\ref{table:parameter} influence the data precision and overall inferencing accuracy.
For area overhead, $\lceil{Bit_{quan}}/{Res_{cell}} \rceil$ is proportional to the number of cells used in the ReRAM-based design to represent the weights, and $Xbar_{size}$ determines the subarray unit size. 
For execution time, note that $Freq$ is inversely proportional to the clock cycle.
% in ReRAM-based designs
The read voltage $V_r$ and the ReRAM cell resistance range~[$R_{on}$,$R_{off}$] affect the ReRAM crossbar energy consumption.

\noindent {\bf MOO formulation.} We formulate the MOO problem for robust inferencing on hardware-efficient ReRAM crossbars with stochastic noise as follows. Our input space consists of two parts: the ReRAM design space and the DNN weights. Let $\mathcal{X}\subseteq \mathcal{R}^{d}$ be the ReRAM design configuration space, which includes the design variables as explained above and also shown in Fig.~\ref{fig:map}.
Each design variable can take values from a bounded candidate set. 
We need a candidate pair consisting of ReRAM design configuration($\mathbf{x} \in \mathcal{X}$) and DNN weights to be able to evaluate all the output objectives. Without loss of generality, we consider maximizing four objective functions: DNN inferencing accuracy, hardware area, execution time, and energy consumption denoted by $f_1(\mathbf{x}), f_2(\mathbf{x}), f_3(\mathbf{x}), f_4(\mathbf{x})$, respectively.
For each candidate ReRAM design, we execute ReSNA to obtain the DNN weights that give rise to maximum accuracy. Subsequently, we evaluate the objective functions $f_{1}(\mathbf{x})$, $f_{2}(\mathbf{x})$, $f_{3}(\mathbf{x})$, $f_{4}(\mathbf{x})$.

A design configuration $\mathbf{x}$ is determined to \textit{Pareto dominate} another design $\mathbf{x'}$ if $f_{i}(\mathbf{x}) \geq f_{i}(\mathbf{x'})\ \forall i$ and there exists some $j \in \{1,2,3,4\}$ such that $f_{j}(\mathbf{x}) > f_{j}(\mathbf{x'})$. 
An optimal solution of a MOO problem is a set of designs $\mathcal{X}^*$ such that no design $\mathbf{x'}\in \mathcal{X}\setminus	\mathcal{X}^*$ pareto-dominates a design $\mathbf{x}\in \mathcal{X}^*$.
The solution set  $\mathcal{X}^*$ is called the \textit{Pareto set}, and the corresponding set of objective function values is called the \textit{Pareto front}. 
Our goal is to achieve a high-quality Pareto front for hardware design while minimizing the total computation cost of function evaluations.

\section{Understanding the Impact of Stochastic Noise}
\label{sec4_stochastic}
In this section, we first discuss the modeling of ReRAM stochastic noise. Next, we demonstrate the impact of stochastic noise on DNN inferencing for specific ReRAM design configurations. Finally, we show that the na\"{\i}ve approach of adding random Gaussian noise cannot improve the robustness of DNN inferencing in the presence of stochastic noise. 
\subsection{Modeling of ReRAM Stochastic Noise}
\label{sec:noise}
{\em Thermal noise} is generated due to the thermal agitation of the charged carriers inside the conductor~\cite{johnson1928thermal}. {\em Shot noise} is an electronic noise that originates from the discrete electrons in the current flow. The thermal and shot noise directly affect the current through a device. We convert the change in current to the equivalent conductance change and model these two noise sources using Gaussian distributions~\cite{feinberg2018making}: $\Delta G_{thermal}=\mathcal{N}(0, {\sqrt{4G\cdot Freq\cdot K_B\cdot T}}/{V})$ and $\Delta G_{shot}=\mathcal{N}(0, {\sqrt{2G\cdot Freq\cdot q\cdot V}}/{V})$, where $G$ and $V$ denote the conductance and terminal voltage respectively.
As shown in Table~\ref{table:parameter}, $Freq$ denotes the operational frequency, and $T$ denotes the temperature. $K_B$ denotes the Boltzmann constant, and $q$ denotes the electron charge.

{\em Random telegraph noise (RTN)} is generated in semiconductors and ultra-thin oxide films.
It can be modeled as a Poisson process~\cite{ielmini2010resistance}, with the parameters for RTN amplitude~($\Delta G_{rtn}$) reported in~\cite{he2019noise}. 

{\em Programming noise} is introduced by the programming variation when values are written to a ReRAM device.
The programming noise can be estimated using a Gaussian distribution with $\Delta G_{prog}=\mathcal{N}(0, \sigma_{prog}G)$ with standard deviation $\sigma_{prog}=0.0658$, according to the experimental study reported in~\cite{lin2019performance}.
\subsection{Impact of ReRAM Stochastic Noise}
\begin{figure}[t]
\centering
\vspace{-6pt}
\includegraphics[width=0.95\linewidth]{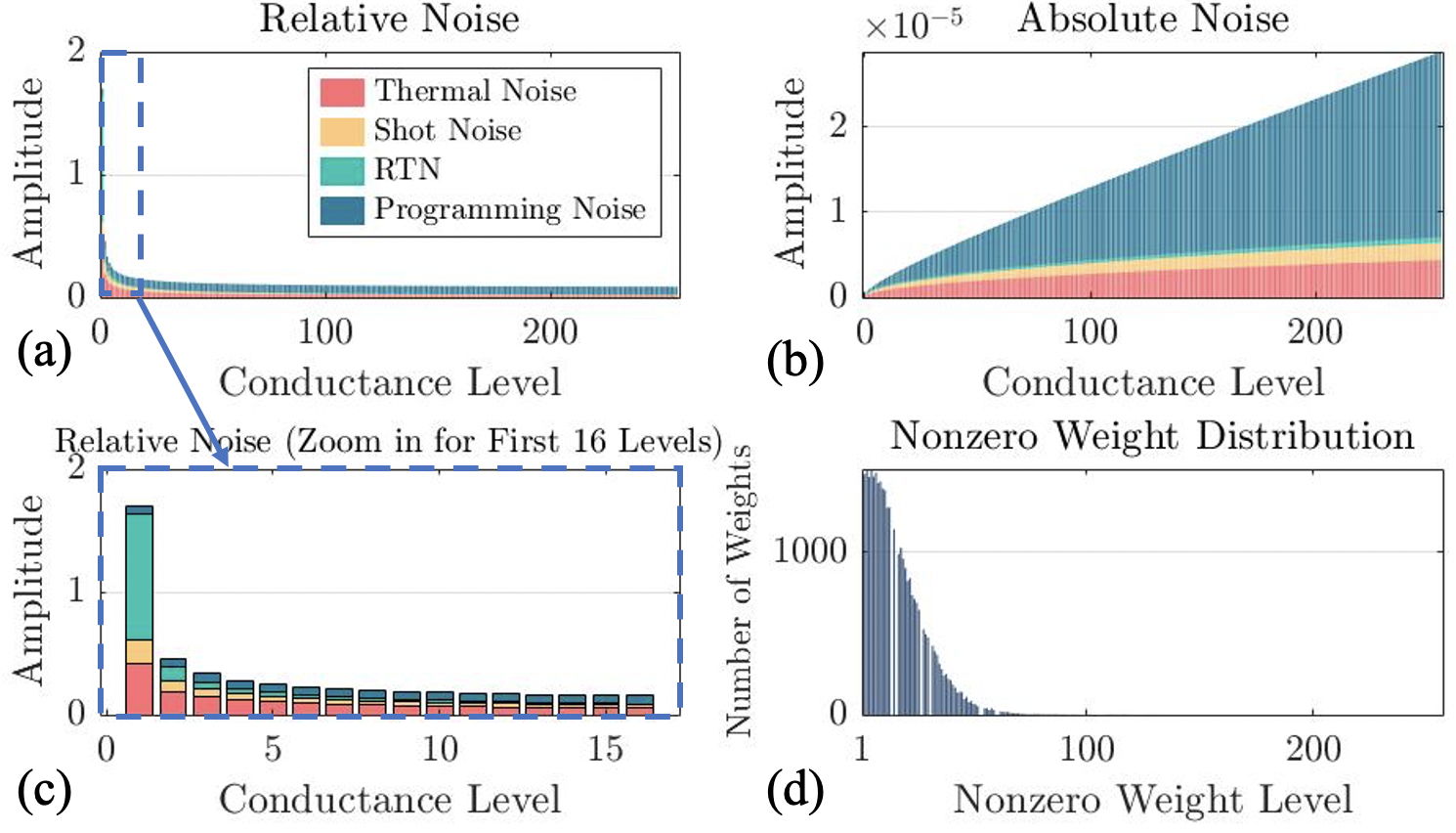}
\vspace{-6pt}
\caption{The distributions of stochastic noise for 8-bit ReRAM cells under $Freq=500\,\mathrm{MHz}$ and $T=350\,\mathrm{K}$: (a) in a relative scale, (b) with absolute values,
(c) zoomed-in version for relative noise,
(d) an example of nonzero weight distribution.}
\label{fig:noise}
\vspace{-18pt}
\end{figure}
%oblivious, 
Fig.~\ref{fig:noise} (a)-(b) show the relative and absolute distributions of four kinds of stochastic noise for 8-bit ReRAM cells with $Freq= 500\,\mathrm{MHz}$ and $T=350\, \mathrm{K}$.
Besides, Fig.~\ref{fig:noise}(c) presents the relative noise for the first 16 levels among the 256 conductance levels.
Relative noise ($\Delta G/G$) measures the noise amplitude divided by the absolute conductance.
Thermal and shot noise have similar patterns: the relative noise is the largest at the smallest conductance level;
it then decreases steadily as $G$ increases.
The relative RTN presents a sharp peak at the first conductance level, while other conductance levels have much smaller relative RTN. 
The programming noise increases with the conductance level in the absolute value.
%and shows a constant impact on relative noise.

We observe from Fig.~\ref{fig:noise}(c) that the overall amplitude is much higher than each individual case, especially when $G$ is small. 
Moreover, it is well known that the trained weights of DNNs are concentrated at small values \cite{seo2019efficient}.
Fig.~\ref{fig:noise}(d) shows an example distribution of nonzero weights, which is extracted from the $\mathrm{19^{th}}$ layer of ResNet20 with 8-bit quantization. 
A large number of the weights in this layer are zero; these are not included in the figure.
Note that at small values of $G$, the first three types of stochastic noise dominate.
Specifically, the thermal and shot noise are sensitive to the operational frequency.
Hence, it is challenging to incorporate high-frequency noise into training.
Lowering the operational frequency and reducing noise amplitude could be an option, as suggested in previous work~\cite{he2019noise}. However, such an approach will seriously constrain the use and potential of ReRAM-based hardware due to the exclusion of high operational frequency and short execution time for DNN inferencing. 

Utilizing high-resolution cells is another challenge.  
For example, increasing the cell resolution from 2 to 8 bits can reduce the number of crossbar arrays by $75\%$ (assuming all other settings are the same), leading to a smaller area and lower latency. However, the noise margin drops by $64\times$ as the cell noise margin is inversely proportional to the number of conductance levels~(i.e., $2^{Res_{cell}}$).
%$2 to the power of cell resolution.
% compared to ReRAM crossbar design with 8-bit cell resolution

\subsection{Performance of Existing Training Approaches}
\begin{figure}[t]
\vspace{-18pt}
\includegraphics[width=0.9\linewidth]{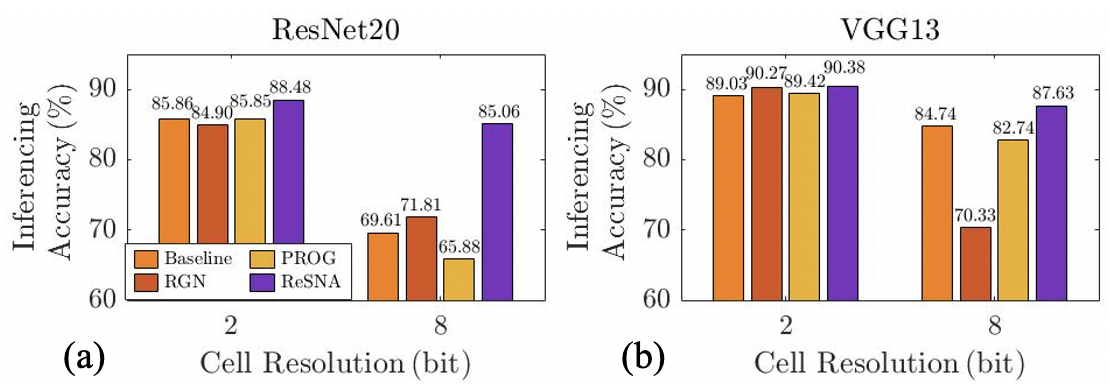}
\vspace{-6pt}
\caption{Results for DNN inferencing on ReRAM crossbars with stochastic noise under different training methods: Baseline, RGN, \textsc{Prog}, and ReSNA training. $Freq=500\,\mathrm{MHz}$, $T=350\,\mathrm{K}$, 128$\times$128 crossbars. (a) ResNet20, (b) VGG13.}
\vspace{-18pt}
\label{fig:compare}
\end{figure}
We next evaluate the DNN inferencing accuracy of several previously proposed ReRAM hardware-aware training methods with ResNet20 and VGG13 and show the results in Fig.~\ref{fig:compare}. 
Weights and activations are quantized to 8 bits.
We consider two different ReRAM cell resolutions, 2 bits and 8 bits.
Models are tested by including the stochastic noise in ReRAM-based hardware.
%, as described in Section~\ref{sec3_problem}
The baseline configuration considers training with no noise. 
RGN denotes a na\"{\i}ve noise-aware approach, which injects random Gaussian noise into the system during training.
The noise standard deviation is related to the maximum absolute weight value, more specifically, $\Delta G_{RGN}=\mathcal{N}(0,\,0.01\cdot \max(G))$. 
\textsc{Prog} considers only the programming noise $\Delta G_{prog}$ with $\sigma_{prog}=0.0658$~\cite{lin2019performance} for training.

We make the following observations from the results shown in Fig.~\ref{fig:compare}. {\bf 1)} The baseline system suffers from stochastic noise, resulting in poor inferencing accuracy. 
{\bf 2)} Previous training methods, such as RGN and \textsc{Prog}, cannot guarantee the mitigation of DNN inferencing accuracy degradation due to stochastic noise. 
This result is mainly due to the mismatch between Gaussian noise and the actual device stochastic noise, as shown in Fig.~\ref{fig:noise}(a). 
{\bf 3)} Increasing the cell resolution from 2 to 8 bits exacerbates the degradation in DNN inferencing accuracy due to the reduced noise margin of the ReRAM devices.

\section{ReSNA: Hardware-Aware Training Approach}
\label{sec5_training}
In this section, we describe the proposed ReRAM-based stochastic-noise-aware (ReSNA) training method that incorporates stochastic noise to improve DNN inferencing accuracy.

We start with a pre-trained model to initialize noise-aware training. 
Batch normalization layers are included after each Conv layer~\cite{joardar2020accured,BO-BN}.
The computation during the training process considers the hardware configurations. 
In each iteration, a new set of emulated device noise is applied in the stochastic noise injection step.
Thus, the loss calculated at the end of the forward pass reflects the influence of stochastic noise. During the backpropagation, gradients of trainable DNN weights are calculated with respect to the loss. 
We keep a copy of the noise-free weight values and perform the gradient updates on this copy. 

The quality of training degrades due to the distortion of the loss function induced by the variation of weight parameters~\cite{chakraborty2018technology}. 
When the variation is large enough, the gradient update during the backpropagation step can be derived from the expected convergence path.
The error due to stochastic noise can propagate and accumulate through the forward path, and similarly, the error in the gradient can propagate and accumulate through the backward path.
Thus, the convergence of the loss function can be affected by the accumulation of the gradient error.

Moreover, our experiments show that Conv layers are less sensitive to device stochastic noise than FC layers. Let $\delta_c$ denote the ReRAM cell's stochastic noise. We assume that one cell is used to represent one weight for simplicity.
During the ReRAM-based FC layer computation involving one weight for a total of $n$ times, the accumulation of the cell's stochastic noise can be approximated as $n^2\delta_c$ (assuming a Gaussian distribution).
As mentioned above, Conv kernels are typically duplicated to improve the throughput of the ReRAM-based hardware.
These copies have identical weights, but the noise can be approximated to have independent statistical distributions.
As the computation involving one weight for $n$ times will be distributed to multiple copies, the accumulation of these cells' stochastic noise can be reduced to $n^2\delta_c/k$.
The value of $k$ is related to the number of duplicate copies as well as the correlations between the stochastic noise of these devices. 
Thus, the device stochastic noise affects FC layers more significantly than Conv layers due to the duplication of Conv kernels. 
To overcome this challenge, we propose two techniques to improve the stability of DNN inferencing accuracy. 
\begin{table}[t]
\vspace{-12pt}
\centering
\caption{DNN inferencing accuracy comparison under various training configurations with ReSNA for the following setting: ResNet20, $Freq=500\,\mathrm{MHz}$, $T=350\,\mathrm{K}$, 8-bit cell resolution, 128$\times$128 crossbars.}
\footnotesize
\vspace{-6pt}
\begin{tabular}{|c||c|c|c} 
\cline{1-3}
\multirow{2}{*}{Training Configuration} & \multicolumn{2}{c|}{ Inferencing Accuracy}            &   \\ 
\cline{2-3}
  & \multicolumn{1}{|c|}{ w/o Voting} & \multicolumn{1}{|c|}{w/ Voting} &   \\ 
  
\cline{1-3}
$\mathrm{Conv_{ideal}+FC_{ideal}}$ &  $69.61\%$ & $73.69\%$ \\ %\hline
$\mathrm{Conv_{500MHz,350K}+FC_{ideal}}$ &  $79.95\%$& $82.61\%$ \\ %\hline
$\mathrm{Conv_{500MHz,350K}+FC_{100MHz,300K}}$ &  $84.05\%$& $85.06\%$  \\ %\hline
$\mathrm{Conv_{500MHz,350K}+FC_{100MHz,350K}}$ &  $81.65\%$& $83.79\%$   \\
$\mathrm{Conv_{500MHz,350K}+FC_{500MHz,350K}}$ &  $10\%$ & $10\%$\\% \hline
\cline{1-3}
\end{tabular}
\label{table:resna}
\vspace{-18pt}
\end{table}
\vspace{-12pt}

\noindent {\bf Applying smaller noise to FC layers.}
Since computing the loss function is critical for the backpropagation step and FC layers are more sensitive to the stochastic noise, we propose to lower the noise level on FC layers for improving the stability of inferencing. Table~\ref{table:resna} compares the performance of ReSNA under different noise configurations. We consider the ResNet20 model with an 8-bit weight and activation quantization, along with a crossbar size of $128\times 128$ in this experiment.
The temperature is set to $350\,\mathrm{K}$, and the operational frequency is set to $500\,\mathrm{MHz}$. 

Table~\ref{table:resna} shows that without including any noise in training ($\mathrm{Conv_{ideal}+FC_{ideal}}$), the DNN inferencing accuracy on this ReRAM design in the presence of stochastic noise is only $69.61\%$.
Including a high-amplitude noise to the entire network ($\mathrm{Conv_{500MHz,350K}+FC_{500MHz,350K}}$) makes the training unstable. 
Applying the device stochastic noise to Conv layers while appropriately reducing the noise level on FC layers (e.g., $\mathrm{Conv_{500MHz,350K}+FC_{100MHz,300K}}$) helps in maintaining the stability of training and improving the DNN inferencing accuracy. 

%\vspace{0.25ex}
\noindent {\bf Majority vote in the classification layer.} Alternatively, FC layers can be duplicated and deployed on different crossbar arrays.
The stochastic noise of a single weight parameter across these copies is not independent but is less correlated than the variations due to accessing the same device multiple times.
We can feed the same input to these duplication layers and take the majority vote (Voting) to determine the predicted output to compensate for the FC layers' impact on the DNN inferencing accuracy. To minimize the area overhead, we apply Voting to only the classification layer (i.e., the last FC layer) with a small number of copies (e.g., 3). The results in Table~\ref{table:resna} demonstrate that this technique can further improve DNN inferencing accuracy, even under the combination of high-amplitude noise and high-resolution cells.

In summary, ReSNA incorporates stochastic noise and enhances stability, leading to better inferencing accuracy than the baseline and previous work, as shown in Fig.~\ref{fig:compare}.

\section{CF-MESMO: Efficient MOO Algorithm}
\label{sec6_moo}
Evaluating DNN inferencing accuracy and hardware efficiency requires execution of the ReSNA training for each ReRAM design configuration; this step is, however, computationally expensive~(e.g., taking over seven hours to execute 100 training epochs for one ReRAM design configuration for ResNet20 with CIFAR10 data). To address this challenge, we propose an efficient information-theoretic MOO algorithm referred to as Continuous-Fidelity Max-value Entropy Search for Multi-objective Optimization (CF-MESMO). Two key innovations here are: first, we formulate continuous-fidelity evaluation of objective functions by varying the number of training epochs of ReSNA. Second, we propose a principled approach to intelligently select the ReRAM configurations and fidelity of ReSNA for evaluation guided by learned statistical models.
% (vary in accuracy of evaluation and computation cost)
\subsection{MOO Formulation with Continuous-Fidelity Evaluations}
For each candidate ReRAM design $x \in \mathcal{X}$, we need to execute the ReSNA method to obtain the DNN weights. Subsequently, we evaluate the objective functions $f_{1}(\mathbf{x})\,$(inferencing accuracy), $f_{2}(\mathbf{x})\,$(hardware area), $f_{3}(\mathbf{x})\,$(execution time), $f_{4}(\mathbf{x})\,$(energy consumption). The cost of evaluation of each ReRAM design configuration can be reduced by making an approximation of the objective function(s). We propose to vary the number of training epochs in ReSNA to trade-off computation cost and accuracy of objective function evaluations (i.e., continuous-fidelity evaluation): small training epochs correspond to lower-fidelity evaluation and vice versa.
Therefore, we formulate this problem as a continuous-fidelity MOO problem where we have access to an alternative function $g_j(\vec{x},z_j)$ for all $j\in\{1,2,3,4\}$. Function $g_j(\vec{x},z_j)$ can make cheaper approximations of $f_j(\vec{x})$ by varying the fidelity variable $z_j \in \mathcal{Z}$.
Without loss of generality, let $\mathcal{Z}=[0, 1]$ be the fidelity space. Fidelities for each function vary in the amount of computational resources consumed and the accuracy of evaluation, where $z_j=0$ and $z_j^*=1$ refer to the lowest and highest fidelity, respectively. At the highest fidelity $z_j^*$, $g_j(\vec{x},z_j^{*})=f_j(\vec{x})$. 
%$f_j$
Let $\mathcal{C}_j(\vec{x},z_j)$ be the cost of evaluating $g_j(\vec{x},z_j)$, i.e., runtime to perform  training using ReSNA for the selected number of training epochs. Evaluation of each ReRAM design configuration $\vec{x}\in \mathcal{X}$ with fidelity vector $\vec{z}=[z_1, z_2, z_3, z_4]$ generates the evaluation vector $\vec{y}\equiv [y_1, y_2, y_3, y_4]$, where $y_j = g_j(\vec{x},z_j)$, and the normalized cost of evaluation is $\mathcal{C}(\vec{x},\vec{z}) = \sum_{j=1}^{4} \left( {\mathcal{C}_j(\vec{x},z_j)}/{\mathcal{C}_j(\vec{x},z_j^*)}\right)$. Our goal is to approximate the Pareto set $\mathcal{X}^*$ by minimizing the overall cost of evaluating candidate ReRAM designs.
\subsection{Overview of CF-MESMO}
CF-MESMO learns a surrogate model using data obtained from past ReRAM design evaluations and then intelligently selects the next candidate ReRAM design and the fidelity of ReSNA pair for evaluation by trading-off exploration with exploitation to quickly direct the search towards Pareto-optimal solutions.
We perform the following steps in each iteration of CF-MESMO as shown in Algorithm 1: {\bf 1)} Select the ReRAM design and fidelity of ReSNA for evaluation that maximizes the information gain per unit cost about the optimal Pareto front based on the current surrogate model. {\bf 2)} Execute the hardware-aware training approach ReSNA to evaluate objective functions with the selected ReRAM design and fidelity pair. 
{\bf 3)} Employ the new training example in the form of ReRAM design configurations (i.e., input) and four objective function evaluations (i.e., output) to update the surrogate model. After convergence is achieved (i.e., Pareto front solution doesn't change in several consecutive iterations), we compute the Pareto front from the aggregate set of objective evaluations and obtain the ReRAM design configurations and DNN weights corresponding to the Pareto front as the resulting solution.
\vspace{-6pt}
\begin{algorithm}[H]
\centering
% \footnotesize
\scriptsize
%\tiny
\caption{CF-MESMO Algorithm}
\textbf{Input}: ReRAM design space $\mathcal{X}$; DNN $\pi$; four objective functions $f_j$ and their continuous approximations $g_j$ using ReSNA training; total cost budget $\mathcal{C}_{total}$.
\vspace{-2ex}
\begin{algorithmic}[1]
\STATE Initialize GP models $\mathcal{GP}_1, \cdots, \mathcal{GP}_4$ via ReRAM design evaluations $D$
\STATE \textbf{While} {$\mathcal{C}_{t} \leq \mathcal{C}_{total} $} \textbf{and} not converged \textbf{do}
 \STATE \quad for each sample $s \in {1,\cdots,S}$: 
 \STATE \quad \quad Sample highest-fidelity functions $\Tilde{f}_j \sim \mathcal{GP}_j(.,z_j^*) $
 \STATE \quad \quad $\mathcal{F}_s^{*} \leftarrow$ Solve {\em cheap} MOO over $(\Tilde{f}_1, \cdots, \Tilde{f}_K)$
\STATE \quad  Select ReRAM design and fidelity pair: \\ \quad $(\vec{x}_{t},\vec{z}_t) \leftarrow \arg max_{\vec{x}\in \mathcal{X},\vec{z}\in \mathcal{Z}} \hspace{2 mm} \alpha_t(\vec{x},\vec{z},\mathcal{F}^{*})$ Equation (\ref{Tappriximation})
\STATE \quad Perform ReSNA training of DNN $\pi$ with ReRAM design and fidelity pair $(\vec{x}_{t},\vec{z}_t)$
\STATE \quad Evaluate objectives $f_1, f_2, f_3, f_4$ for trained DNN on ReRAM design $\vec{x}_{t}$
\STATE \quad Update the total cost: $\mathcal{C}_t \leftarrow \mathcal{C}_t + \mathcal{C}(\vec{x}_t,\vec{z}_t)$
\STATE \quad Aggregate training data: $\mathcal{D} \leftarrow \mathcal{D} \cup \{(\vec{x}_{t}, \vec{y}_{t},\vec{z}_t)\}$ 
\STATE \quad Update surrogate statistical models $\mathcal{GP}_1,\cdots, \mathcal{GP}_4$ 
\STATE \quad $t \leftarrow t+1$
\STATE \textbf{end}
\STATE \textbf{return} Pareto set and Pareto front of objective functions $f_1(x), \cdots,f_4(x)$
\end{algorithmic}
\label{alg:CFMESMO}
\end{algorithm}
\vspace{-12pt}
%\vspace{0.25ex}
\noindent {\bf Surrogate models for continuous-fidelity.} 
%We learn surrogate statistical models from the past ReRAM design evaluation data. 
Surrogate models guide the selection of candidate ReRAM designs to quickly uncover high-quality Pareto fronts. Our training data $\mathcal{D}$ for surrogate models after $t$ iterations consists of $t$ training examples of input-output pairs.  
We employ Gaussian processes
(GPs)~\cite{williams2006gaussian} as our choice of the statistical model due to their superior uncertainty quantification ability.
We learn four surrogate statistical models $\mathcal{GP}_1,\cdots,\mathcal{GP}_4$ from $\mathcal{D}$, where each model $\mathcal{GP}_j$ corresponds to the $j$th function $g_j$. Continuous-fidelity GPs (CF-GPs) are capable of modeling functions with continuous fidelities within a single model. Hence, we employ CF-GPs to build surrogate statistical models for each function \cite{kandasamy2017multi}. 
A CF-GP is a random process defined over the input space and the fidelity space, characterized by a mean function $\mu: \mathcal{X} \times \mathcal{Z} \rightarrow \mathbb{R} $ and a covariance or kernel function $\kappa: (\mathcal{X} \times \mathcal{Z})^2 \rightarrow \mathbb{R} $. 
We denote the posterior mean and standard deviation of $g_j$ by $\mu_{g_j}(\vec{x},z_j)$ and $\sigma_{g_j}(\vec{x},z_j)$.
We denote the posterior mean and standard deviation of the highest fidelity functions $f_j(\vec{x})=g_j(\vec{x},z_j^*)$ by $\mu_{f_j}(\vec{x})=\mu_{g_j}(\vec{x},z_j^*)$ and $\sigma_{f_j}(\vec{x})=\sigma_{g_j}(\vec{x},z_j^*)$, respectively.
\subsection{Selecting ReRAM Design to Evaluate via Information Gain} 

The effectiveness of CF-MESMO critically depends on the reasoning mechanism to select the candidate ReRAM design and fidelity of ReSNA pair for evaluation in each iteration. Therefore, we propose an information-theoretic approach to perform this selection.
{\em The key idea is to find the ReRAM design and fidelity pair $\{\vec{x}_t, \vec{z}_t\}$ that maximizes the information gain ($I$) per unit cost about the Pareto front of the highest fidelities} (denoted by $\mathcal{F}^*$), where $\{\vec{x}_t, \vec{z}_t\}$ represents a candidate ReRAM design configuration $\vec{x}_t$ evaluated at fidelities $\vec{z}_t$ at iteration $t$. CF-MESMO performs the joint search over the input space $\mathcal{X}$ and the fidelity space $\mathcal{Z}$:
% = $[z_1, z_2, \cdots, z_4]$
%\vspace{0.75ex}
\begin{align}
    (\vec{x}_{t},\vec{z}_t) \leftarrow \argmax_{\vec{x}\in \mathcal{X},\vec{z}\in \mathcal{Z}}\hspace{2 mm}\alpha_t(\vec{x},\vec{z}), ~
\end{align}
\begin{align}
          \text{where} \quad \alpha_t(\vec{x},\vec{z}) &= I(\{\vec{x}, \vec{y},\vec{z}\}, \mathcal{F}^* | \mathcal{D}) / \mathcal{C}(\vec{x},\vec{z}). \label{af:def}
\end{align}

In this joint search, the computation cost~$\mathcal{C}(\vec{x},\vec{z})$ is considered in Equation~(\ref{af:def}). The information gain in Equation (\ref{af:def}) is the expected reduction in entropy $H(.)$ of the posterior distribution $P(\mathcal{F}^* | \mathcal{D})$ due to the evaluation of the ReRAM design $\vec{x}$ at fidelity vector $\vec{z}$. According to the symmetric property, the information gain can be rewritten as follows:
%\vspace{0.75ex}
\begin{align}
    I(\{\vec{x}, \vec{y},\vec{z}\}, \mathcal{F}^{*} | \mathcal{D}) &= H(\vec{y} | \mathcal{D}, \vec{x},\vec{z}) - \mathbb{E}_{\mathcal{F}^{*}} [H(\vec{y} | \mathcal{D},   \vec{x},\vec{z}, \mathcal{F}^{*})]. \numberthis \label{eqn_symmetric_ig}
\end{align}
The first term in Equation (\ref{eqn_symmetric_ig}) is the entropy of a four-dimensional Gaussian distribution that can be computed as follows:
\begin{align}
H(\vec{y} | \mathcal{D}, \vec{x},\vec{z}) = \sum_{j = 1}^4 \ln (\sqrt{2\pi e} ~ \sigma_{g_j}(\vec{x},z_j)). \numberthis \label{firstpart}
\end{align}
The second term in Equation (\ref{eqn_symmetric_ig}) is an expectation over $\mathcal{F}^{*}$ and can be approximated using Monte-Carlo sampling: 
\begin{align}
    \mathbb{E}_{\mathcal{F}^{*}} [H(\vec{y} | \mathcal{D},   \vec{x},\vec{z}, \mathcal{F}^{*})] \simeq \frac{1}{S} \sum_{s = 1}^S [H(\vec{y} | \mathcal{D},   \vec{x},\vec{z}, \mathcal{F}^{*}_s)], \label{eqn_summation}
\end{align}
where $S$ denotes the number of samples, and $\mathcal{F}^{*}_s$ denotes a sample Pareto front achieved over the highest fidelity functions sampled from the surrogate models.  
To solve Equation (\ref{eqn_summation}), we provide solutions to construct Pareto front samples $\mathcal{F}^{*}_s$ and to compute the entropy of a given Pareto front sample $\mathcal{F}^{*}_s$. 

\vspace{0.25ex}

\noindent {\bf Computation of Pareto front samples:} We sample the highest fidelity functions $\Tilde{f}_1,\cdots,\Tilde{f}_4$ from the posterior CF-GP models.
Then, we solve a cheap MOO problem over the sampled functions with the NSGA-II algorithm \cite{deb2002fast} and compute the sample Pareto front $\mathcal{F}^{*}_s$.
%four popular
\vspace{0.25ex}

\noindent {\bf Entropy computation for a given Pareto front sample:} Let $\mathcal{F}^{*}_s = \{\vec{v}^1, \cdots, \vec{v}^l \}$ be the sample Pareto front, where $l$ denotes the size of the Pareto front and each element  $\vec{v}^i = \{v_1^i,\cdots,v_4^i\}$ is evaluated at the sampled highest-fidelity function. The following inequality holds for each component $y_j$ of $\vec{y}$ in the entropy term $H(\vec{y} | \mathcal{D},   \vec{x},\vec{z}, \mathcal{F}^{*}_s)$:
\begin{align}
 y_j &\leq f_s^{j*} \quad \forall j \in \{1,\cdots,4\}, \label{inequality}
\end{align}
where $f_s^{j*} = \max \{v_j^1, \cdots v_j^l \}$.
Essentially, this inequality means that the $j^{th}$ component of $\vec{y}$ is upper-bounded by the maximum of $j^{th}$ components of sample Pareto front $\mathcal{F}^{*}_s$.
%  a value, which is all $l$ vectors $\{\vec{v}^1, \cdots, \vec{v}^l \}$ in the 

The proof of Equation (\ref{inequality}) falls in two cases\footnote{For ease of notation, we drop the dependency on $\vec{x}$ and $\vec{z}$. 
We use $f_j$ to denote $f_j(\vec{x})=g_j(\vec{x},z_j^*)$ the evaluation of the highest fidelity $z_j^*$ and $y_j$ to denote $g_j(\vec{x},z_j)$ the evaluation of $g_j$ at a lower fidelity $z_j \neq z_j^*$.}:
{\bf a)} If $y_j$ is evaluated at the highest fidelity (i.e, $z_j=z_j^*$ and $y_j=f_j$), we prove by contradiction.
% for inequality in Equation (\ref{inequality})
Suppose there exists some component $f_j$ of $\vec{f}$ such that $f_j > f_s^{j*}$.
However, by definition, since no point dominates $\vec{f}$ in the $j$th dimension, $\vec{f}$ is a non-dominated point.
This results in $\vec{f} \in \mathcal{F}^*_s$, which is a contradiction.
Thus, Equation (\ref{inequality}) holds.
{\bf b)} If $y_j$ is evaluated at one of the lower fidelities (i.e, $z_j \neq z_j^*$), we refer to the assumption that the value of an objective evaluated at lower fidelity is smaller than that evaluated at higher fidelity, i.e., $y_j \leq f_j \leq f_s^{j*}$. This assumption is true in our problem setting, where the DNN inferencing accuracy improves with more training epochs of ReSNA.

Following Equation (\ref{inequality}) and the independence of CF-GP models, we further decompose the entropy of a set of independent variables according to the entropy measure property~\cite{information_theory}:
\begin{align}
H(\vec{y} | \mathcal{D},   \vec{x},\vec{z}, \mathcal{F}^{*}_s) \simeq \sum_{j=1}^4 H(y_j|\mathcal{D}, \vec{x},z_j,f_s^{j*}). \label{eqn_sep_ineq}
% \vspace{-1.5ex}
\end{align} 
Equation (\ref{eqn_sep_ineq}) requires the entropy computation of $p(y_j|\mathcal{D}, \vec{x},z_j,f_s^{j*})$. This conditional distribution can be expressed as $H(y_j|\mathcal{D}, \vec{x},z_j,y_j\leq f_s^{j*})$.
%depends on the value of $z_j$ and
As Equation (\ref{inequality}) states that $y_j\leq f_s^{j*}$ holds under all fidelities, the entropy of $p(y_j|\mathcal{D}, \vec{x},z_j,f_s^{j*})$ can be approximated by the entropy of a truncated Gaussian distribution as:
%  \vspace{-2ex}
\begin{align}
H(y_j|\mathcal{D}, \vec{x},z_j,y_j\leq f_s^{j*})= & \ln(\sqrt{2\pi e} ~ \sigma_{g_j}) +  \ln \Phi(\gamma_s^{(g_j)}) \nonumber \\ & - \frac{\gamma_s^{(g_j)} \phi(\gamma_s^{(g_j)})}{2\Phi(\gamma_s^{(g_j)})},   \numberthis \label{entropyapprox1}
\end{align}
% \vspace{-1ex}
where $\gamma_s^{(g_j)} = \frac{f_s^{j*} - \mu_{g_j}}{\sigma_{g_j}}$. 
Functions $\phi$ and $\Phi$ are the probability density and cumulative distribution function of the standard normal distribution, respectively.
%This entropy can be computed using the following approximation.
From Equations (\ref{firstpart}), (\ref{eqn_summation}), and (\ref{entropyapprox1}), we get the expression as shown below:
%  \vspace{-2.5ex}
\begin{align}
    \alpha_t(\vec{x},\vec{z},\mathcal{F}^{*})=&\frac{1}{\mathcal{C}(\vec{x},\vec{z}) S}\sum_{j=1}^4 \sum_{s=1}^S  \frac{\gamma_s^{(g_j)}\phi(\gamma_s^{(g_j)})}{2\Phi(\gamma_s^{(g_j)})} - \ln(\Phi(\gamma_s^{(g_j)})). \numberthis \label{Tappriximation}
\end{align}
Therefore, in Algorithm~\ref{alg:CFMESMO}, we select the next ReRAM design and the fidelity of ReSNA pair that maximizes the information gain per unit cost about the optimal Pareto front based on Equation~(\ref{Tappriximation}).

\section{Experiments and Results}
\label{sec7_experiment}
In this section, we first explain the details of the experimental setup.
Next, we evaluate the effectiveness of ReSNA in improving the inferencing accuracy.
Finally, we show that CF-MESMO can achieve high-quality Pareto fronts for DNN inferencing on ReRAM crossbars and analyze the Pareto sets for different DNN models.
\subsection{Experimental Setup}
\label{sec7_1:experiments}
%\vspace{-3pt}
\begin{table}[!b]
\vspace{-12pt}
\caption{Experiments setup details.}
\label{table:configuration}
\vspace{-4pt}
\footnotesize
\centering
(a) Network configurations.\\
\vspace{2pt}
\begin{tabular}{|c|c|c|c|}
\hline
 Network                       & \# of channels in Conv layers                                                              & \begin{tabular}[c]{@{}c@{}}Unquantized \\Accuracy \end{tabular} &\begin{tabular}[c]{@{}c@{}}Quantized \\Accuracy \end{tabular}              \\ \hline
 ResNet20       &\begin{tabular}[c]{@{}c@{}}16, {[}16,16{]}$\times$3, {[}32, 32{]}$\times$3,\\ {[}64, 64{]}$\times$3   \end{tabular}                                     & 91.65\% &89.61\%    \\             
 ResNet32 &\begin{tabular}[c]{@{}c@{}}16, {[}16,16{]}$\times$5, {[}32, 32{]}$\times$5,\\ {[}64, 64{]}$\times$5   \end{tabular}                                   & 92.81\%& 90.06\%            \\
 ResNet44 & 
\begin{tabular}[c]{@{}c@{}}16, {[}16,16{]}$\times$7, {[}32, 32{]}$\times$7,\\ {[}64, 64{]}$\times$7   \end{tabular}                                      & 93.24\% &91.54\%              \\  
 VGG11& \begin{tabular}[c]{@{}c@{}}64, 128, 256, 256, 512, 512, \\512, 512    \end{tabular}                                                 & 92.18\% &88.07\%              \\ 
  VGG13                        & \begin{tabular}[c]{@{}c@{}}64, 64, 128, 128, 256, 256,\\  512, 512, 512, 512\end{tabular} & 93.64\% & 91.14\%                  \\ \hline
 \begin{tabular}[c]{@{}c@{}}{ResNet18}\\{\tiny for CIFAR-100}\end{tabular} & \begin{tabular}[c]{@{}c@{}}64, {[}64, 64{]}$\times$2, {[}128,128{]}$\times$2,\\
{[}256,256{]}$\times$2, {[}512,512{]}$\times$2 \end{tabular}  &74.57\% &  71.92\%\\ \hline
\end{tabular}
\label{table:network}
\vspace{2pt}

(b) ReRAM parameters in ReSNA.\\
\vspace{2pt}
\begin{tabular}{|c||c|}
\hline
 Parameter & Value \\ \hline
$Bit_{quan}$ & $8$ bit       \\ %\hline
$R_{on},R_{off}$ & $3.03\,\mathrm{k\Omega}$, $3.03\,\mathrm{M\Omega}$       \\ %\hline

$Res_{DAC},Res_{ADC}$ & $8$ bits, $8$ bits      \\
$V_{r}$ & $\mathrm{1.65V}$      \\
$\sigma_{prog}$ & $0.0658$      \\ %\hline
\hline
\end{tabular}
%\label{table:parameter_number}
\vspace{2pt}\\

(c) Design space configurations.\\
\vspace{2pt}
\begin{tabular}{|c||c|}
\hline
 Parameter & Candidate values \\ \hline
$Res_{cell}$ & $8/4/3/2/1$-bit      \\ %\hline
$Freq$ & $10\,\mathrm{MHz}$-$1000\,\mathrm{MHz}$     \\
$T$ & $300\,\mathrm{K}$-$400\,\mathrm{K}$     \\
$Xbar_{size}$ & $32\times32$, $64\times64$, $128\times128$  \\
\hline
\end{tabular}
\end{table}

We evaluate ReSNA with five different DNNs---ResNet20, ResNet32,  ResNet44~\cite{he2016deep}, VGG11, and VGG13~\cite{simonyan2014very} on the CIFAR-10 dataset~\cite{krizhevsky2009learning}. 
The CIFAR-10 dataset contains $50,000$ training images and $10,000$ testing images, which belong to $10$ classes. 
Furthermore, to validate the scalability of our method, we also evaluate the performance of ResNet18~\cite{he2016deep} using the CIFAR-100 dataset~\cite{krizhevsky2009learning}. The number of training and testing images in CIFAR-100 is the same as in CIFAR-10, but these images belong to $100$ classes.
The image size is $28\times28\times3$, and the training and testing batch size is $64$.
Table~\ref{table:configuration}(a) summarizes deep neural network configurations, including the numbers of channels in Conv layers, the inferencing accuracy with unquantized weights and activations, and the inferencing accuracy for 8-bit weights and activations. Note that testing on diverse DNNs is more important to test the effectiveness of our approach. Hence, due to space constraints, we provide results on limited datasets noting that our methodology and findings are general.
%when quantizing weights and activations to 8 bits.

We implement the ReSNA method on ReRAM crossbars with stochastic noise using the PytorX simulator~\cite{he2019noise}.  
ReSNA uses stochastic gradient descent with a learning rate of $0.001$ and a momentum of $0.9$. The maximum number of training epochs is $100$.
For ResNet18, we decay the learning rate by $0.2$ for every 20 epochs.
%~(SGD)
Each inferencing test consists of $10$ independent runs with stochastic noise, and we report the average inferencing accuracy.
All the training and inferencing are conducted on NVIDIA Titan RTX GPU with a memory of $24\,\mathrm{GB}$ and a memory bandwidth of $672\,\mathrm{GB/s}$.

Table~\ref{table:configuration}(b) summarizes the ReRAM device parameters.
The cell resolution,  operational frequency, temperature, and crossbar size are inputs to the MOO framework.
Table~\ref{table:configuration}(c) shows the allowable input configurations.
We use NeruoSim~\cite{peng2019dnn+} along with the $32\,\mathrm{nm}$ technology node parameters to evaluate the hardware area, execution time, and energy consumption.
The ReRAM crossbar and peripheral configurations are adopted from~\cite{peng2019optimizing}.
For ReSNA with Voting, we duplicate the kernels of the classification layer.
For the MOO problem, we run CF-MESMO for a maximum of $100$ iterations with $10$ available fidelity selections.
The baselines for evaluating the efficiency of CF-MESMO are the random search and NSGA-II. We utilize the NSGA-II implementation from \textit{Platypus} python library.

\subsection{ReSNA Results}
\begin{figure}[t]
\vspace{-12pt}
\centering
\includegraphics[width=\linewidth]{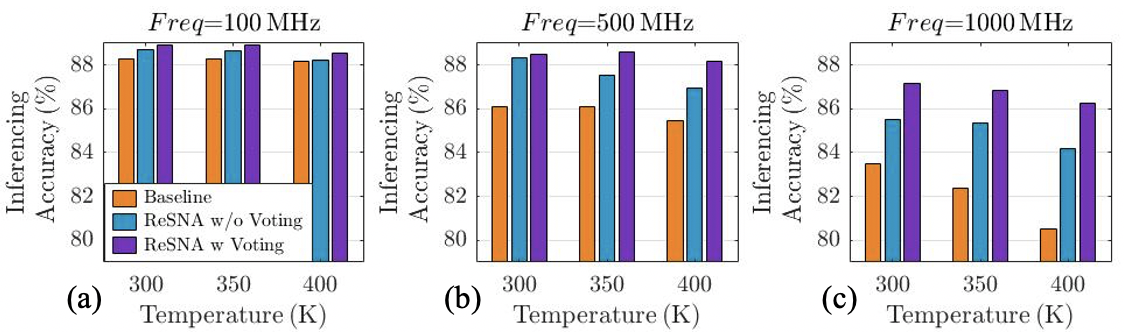}
\vspace{-12pt}
\caption{ReRAM inferencing accuracy under various temperature and frequency settings. 8-bit cell resolution, 64$\times$64 crossbars. ResNet20 on the CIFAR-10 dataset.}
\vspace{-18pt}
\label{fig:resna_acc}
\end{figure}
\noindent{\bf Inferencing accuracy.}
We show representative results for ReSNA inferencing accuracy
with multiple temperature and frequency settings with respect to the baseline.
Recall that the baseline configuration considers training with no noise and performs inferencing in the presence of stochastic noise. 
Fig.~\ref{fig:resna_acc}(a)-(c) show the inferencing accuracy for ResNet20 with 8-bit cell resolution and 64$\times$64 crossbars.
On average, ReSNA without Voting outperforms the baseline by 1.62$\%$ over all the test conditions. ReSNA with Voting increases the overall inferencing accuracy by 2.57$\%$.
Considering the extreme design configuration, i.e., at 1000$\,\mathrm{MHz}$ and 400$\,\mathrm{K}$, ReSNA with Voting outperforms the baseline by 5.47$\%$. Note that this extreme case assumes that in the future, ReRAM-based hardware will run at this high frequency. We also validate the performance of the ReSNA method with the ResNet18 on the CIFAR-100.
%, as shown in Fig.~\ref{fig:resna_acc}(d)
With a frequency of 500$\,\mathrm{MHz}$ and a temperature of 350$\,\mathrm{K}$, ReSNA achieves $70.80\%$ inferencing accuracy compared with the baseline inferencing accuracy of $69.37\%$.
With a frequency of 1000$\,\mathrm{MHz}$ and a temperature of 350$\,\mathrm{K}$, ReSNA achieves $68.23\%$ inferencing accuracy compared with the baseline inferencing accuracy of $64.45\%$.

In summary, ReSNA can achieve considerable inferencing accuracy improvement under stochastic noise with different ReRAM design configurations, DNNs, and datasets.

\noindent{\bf Design trade-offs considering different objectives.}
\label{sec7_2_result_acc}
\begin{figure}[t]
\vspace{-12pt}
\centering
\includegraphics[width=0.88\linewidth]{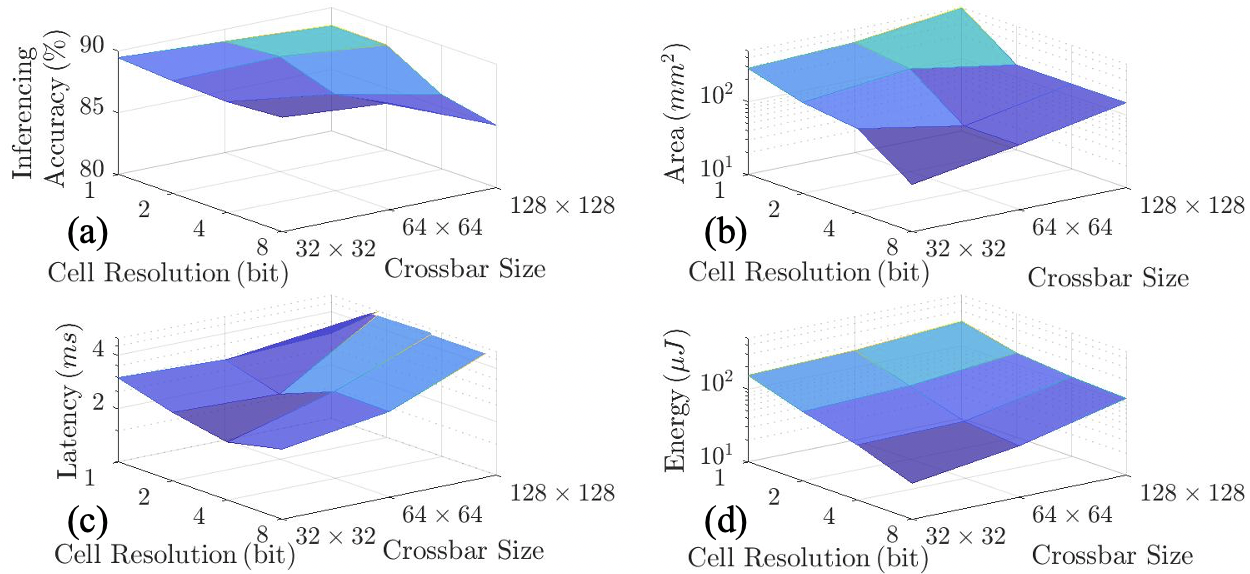}
\vspace{-6pt}
\caption{ReRAM-based design trade-offs under different cell resolution and crossbar size settings. ResNet20, $Freq=500\,\mathrm{MHz}$, $T=350\,\mathrm{K}$. (a) inferencing accuracy, (b) hardware area, (c) latency, (d) energy consumption.}
\vspace{-18pt}
\label{fig:resna_tradeoff}
\end{figure}
We further explore the design trade-offs considering the inferencing accuracy, hardware area, execution time, and energy consumption.
As we have explored the effects of frequency and temperature on inferencing accuracy in the previous analysis, we focus on cell resolution and crossbar size in this discussion.
Fig.~\ref{fig:resna_tradeoff} shows the impact of cell resolution and the crossbar size on ResNet20 at 500$\,\mathrm{MHz}$ and 350$\,\mathrm{K}$.

Fig.~\ref{fig:resna_tradeoff}(a) shows that inferencing accuracy using the ReSNA method. When other settings remain the same, the inferencing accuracy steadily increases as the cell resolution reduces due to the improved noise margin.
As discussed in Section~\ref{sec3_problem}, the ReRAM crossbar array outputs are accumulated across the columns, and hence the crossbar size affects the noise accumulation. 
Therefore, a large crossbar with high cell resolution is not an optimal design choice with this setting from the inferencing accuracy perspective.

From the area perspective, the 32$\times$32 crossbar with 8-bit cell resolution is the best, while 64$\times$64 crossbar has the least area when the cell resolution is 4-bit. 
The area evaluation reveals a high correlation between the cell resolution and crossbar size. Fig.~\ref{fig:resna_tradeoff}(c) shows that the minimum latency is achieved by the 64$\times$64 crossbar with the cell resolution of 2-bit, though this particular configuration incurs a relatively larger area overhead than the configuration with 4-bit cell resolution.
Fig.~\ref{fig:resna_tradeoff}(d) shows that the energy consumption with large cell resolution and small crossbar size is relatively modest.

\subsection{Results on Using CF-MESMO to Optimize ReRAM Crossbars}

Fig.~\ref{fig:resna_tradeoff} also indicates that different objectives have different optimal design configurations, and a global optimal design configuration is not achievable.
Note that the operational frequency and temperature considered above are discrete data points selected for initial performance evaluation. However, the temperature can take any value from $300\,\mathrm{K}$ to $400\,\mathrm{K}$.
Assuming the temperature resolution to be $0.1\,\mathrm{K}$, we can get $1,000$ data points.
Similarly, by considering other inputs, we can estimate the number of all design configurations to be $1.485\times 10^7$.
While any MOO framework can search over this enormous space to achieve the Pareto front to establish the suitable design trade-offs, the computation cost associated with this search is prohibitively high (e.g., it takes nearly $30$ GPU days to run the ReSNA training on the PytorX simulator~\cite{he2019noise} for 100 configurations).
In contrast, the proposed CF-MESMO framework does not traverse through all the data points but can achieve a high-quality Pareto front with significantly reduced computation cost.
We use the hypervolume, which measures the volume between the Pareto front and a reference point~\cite{zitzler1999multiobjective}, to indicate the quality of the Pareto front. 

\noindent{\bf CF-MESMO vs. NSGA-II and random search.}
\begin{figure}[t]
\centering
\vspace{-12pt}
\includegraphics[width=0.95\linewidth]{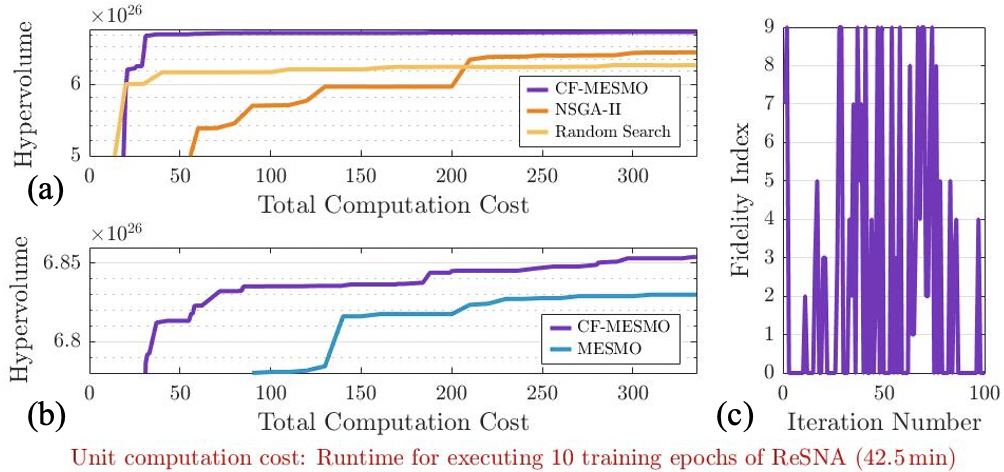}
\vspace{-3pt}
\caption{CF-MESMO framework results for ResNet20: (a) CF-MESMO hypervolume result compared with NSGA-II and random search,
(b) CF-MESMO and MESMO hypervolume results,
(c) fidelity index over the CF-MESMO iterations.}
\vspace{-15pt}
\label{fig:moo_convergence}
\end{figure}
Fig.~\ref{fig:moo_convergence}(a) illustrates the hypervolume result of CF-MESMO compared with NSGA-II and random search under the same computation cost. The unit cost is defined as the runtime for executing 10 training epochs of the ReSNA method~(e.g., $42.5$ minutes based on our setting for ResNet20).
We observe that {\bf 1)} CF-MESMO can achieve a higher-quality Pareto optimal set for the same total computation cost for ReRAM design evaluation;
{\bf 2)} CF-MESMO can produce higher quality Pareto front using significantly lower cost compared to NSGA-II and random search;
{\bf 3)} CF-MESMO achieves $90.91\%$ and $91.21\%$ reduction in computation cost to reach the same quality Pareto front as NSGA-II and random search, respectively.

\noindent{\bf Continuous fidelity vs. single maximum fidelity.}
We utilize a continuous fidelity setting in CF-MESMO.
As a comparison, we run our optimization method with the single maximum fidelity setting (100 training epochs of ReSNA to evaluate each ReRAM design), denoted as MESMO.
Fig.~\ref{fig:moo_convergence}(b) compares the hypervolume of these two methods considering the computational cost. 
The continuous-fidelity setting in CF-MESMO can guarantee higher quality Pareto front with lower computation cost when compared to the single maximum fidelity algorithm MESMO.
Specifically, CF-MESMO lowers the computation cost by $78.18\%$ to reach the same quality Pareto front as MESMO.
Note that both CF-MESMO and MESMO can outperform NSGA-II and random search, as we observe from Fig.~\ref{fig:moo_convergence}(a)-(b).
These results validate the effectiveness of the proposed CF-MESMO algorithm for ReRAM based MOO optimization.

Fig.~\ref{fig:moo_convergence}(c) shows the fidelity index (small index means ReRAM design evaluation using ReSNA with a small number of training iterations) selection over iterations in CF-MESMO.
The optimization starts with a large fidelity index, e.g., 7 and 9 at the $1^{st}$ and $2^{nd}$ iterations in the initialization.
During the $3^{rd}$ to $25^{th}$ iterations, a small fidelity index is preferred to get a fast approximation of the Pareto front by identifying the promising areas of the ReRAM design space. 
After that, a larger fidelity index is selected to approach the optimal Pareto front by evaluating candidates from this set of promising ReRAM designs.
By using this continuous-fidelity setting, we can exclude the non-promising configurations in an early stage. As shown in Equation~(\ref{af:def}), high fidelity is only utilized when the predicted information gain per unit cost is large.

\vspace{0.25ex}
\noindent{\bf Optimization results for different DNNs.}
\begin{figure}[t]
\centering
\includegraphics[width=\linewidth]{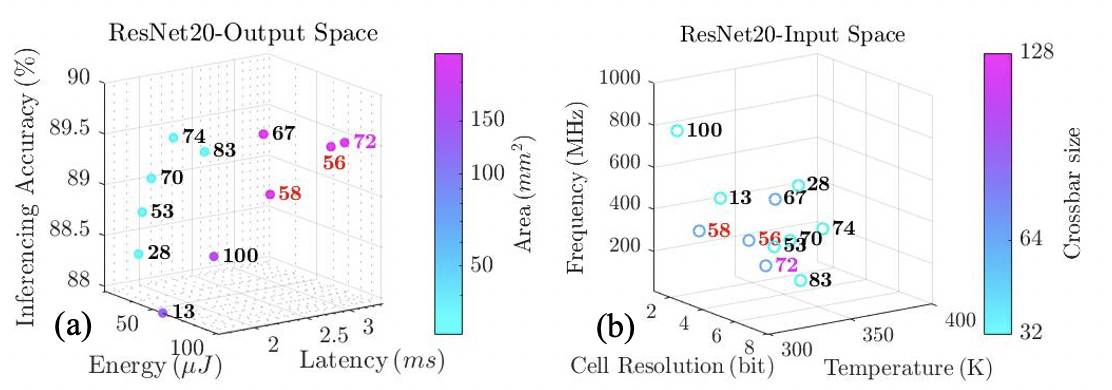}
\vspace{-15pt}
\caption{CF-MESMO framework Pareto front and Pareto set for ResNet20 in a 3-dimensional plot with colorbar, the iteration number is labeled next to each data point: (a) output space, (b) input space.}
\vspace{-18pt}
\label{fig:result_vis_1}
\end{figure}
\begin{figure}[t]
\vspace{-12pt}
\centering
\includegraphics[width=\linewidth]{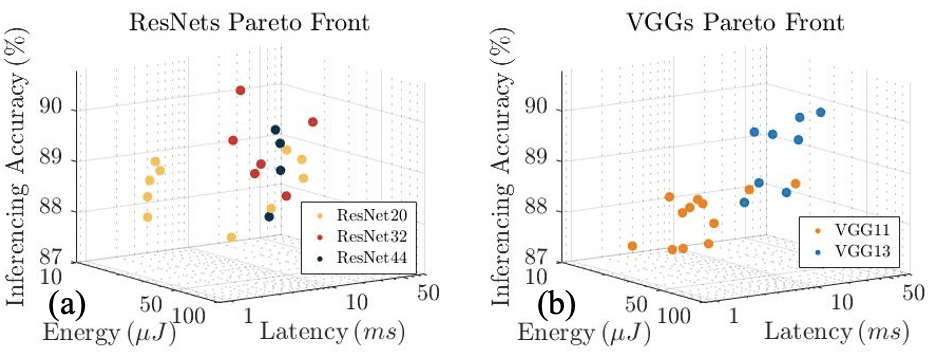}
\vspace{-15pt}
\caption{CF-MESMO framework Pareto front result for ResNets and VGGs in a 3-dimensional plot: (a) ResNets, (b) VGGs.}
\vspace{-18pt}
\label{fig:result_vis}
\end{figure}
We use the proposed CF-MESMO framework to achieve robust DNN inferencing with an efficient ReRAM-based hardware platform.
For the ResNet20 network, CF-MESMO obtains $11$ Pareto optimal designs over $100$ iterations. 
Fig.~\ref{fig:result_vis_1}(a) represents these designs in the output space, including the inferencing accuracy, latency, energy consumption, and area overhead. 
Each data point is labeled with the corresponding iteration number. 
Although all the $11$ design instances appear at the Pareto front, some of them can be excluded due to efficiency constraints.
For instance, designs `56' and `58' consume $113.3\%$ and $105.2\%$ more energy compared to the average of the other designs, while
design `72' requires $57.4\%$ more execution time compared to the average of the other designs.
We mark the high-energy design points with red and the high-latency design point with magenta in Fig.~\ref{fig:result_vis_1}.
According to the input space shown in Fig.~\ref{fig:result_vis_1}(b), a relatively low temperature~($\mathrm{300K}$-$\mathrm{350K}$) and small and medium crossbar sizes~(32$\times$32 and 64$\times$64) are recommended for ResNet20.

Fig.~\ref{fig:result_vis} shows the Pareto fronts for ResNets and VGGs. 
Note that the area evaluation dimension is not included in the plot for ease of illustration, and latency is normalized under the same area constraint.
It should be noted that ResNet32 and VGG13 achieve the best inferencing accuracy for each network class. 
Comparing Fig.~\ref{fig:result_vis}(a) with Fig.~\ref{fig:result_vis}(b), we see that the Pareto fronts for ResNet and VGG have different distributions, while there is a large overlap among the various clusters within the same network class. e.g., the three clusters in Fig.~\ref{fig:result_vis}(a). 
These results imply that the network structure is not the only factor to determine hardware efficiency. 
In designing ReRAM-based accelerators, we should first set the expected inferencing accuracy and hardware efficiency target and then choose the network using the Pareto front.

Based on the Pareto set results from all the evaluations, we make the following observations:
{\bf 1)} Designs with high temperature can appear in the Pareto front, but the number of those cases is low. 
{\bf 2)} For small DNN models, the channel number is relatively small. Large-sized crossbar results in a low utilization rate and thus is not an optimal choice.
As the model size increases, a large-sized crossbar becomes a preferred choice.
{\bf 3)} As ReSNA improves the inferencing accuracy, high ReRAM cell resolution and high frequency become the preferred candidates for robust DNN inferencing.

\section{Conclusions}
We have presented a ReRAM-based accelerator design and optimization framework to achieve robust DNN inferencing in the presence of stochastic noise. The efficiency of this framework depends on uncovering the Pareto-optimal ReRAM designs to establish a suitable trade-off considering inferencing accuracy, area overhead, execution time, and energy consumption. We have solved this challenging multi-objective optimization (MOO) problem by introducing a Continuous-Fidelity Max-value Entropy Search-based MOO framework, called CF-MESMO. CF-MESMO is aided by a hardware-aware training method to handle stochastic noise, called ReSNA. The CF-MESMO framework provides a high-quality Pareto front for robust DNN inferencing on hardware-efficient ReRAM crossbars with stochastic noise. On average, ReSNA achieves $2.57\%$ inferencing accuracy improvement for ResNet20 on the CIFAR-10 dataset with respect to the baseline configuration. Moreover, the CF-MESMO framework achieves $90.91\%$ reduction in computation cost compared with the popular MOO framework NSGA-II to reach the same quality Pareto front as NSGA-II.
\label{sec8_conclusion}

\footnotesize
\bibliographystyle{ieeetr}
\bibliography{main}

\end{document}